\documentclass[aps,ams,onecolumn,nofootinbib,showpacs ]{revtex4}

\usepackage{graphicx}\usepackage{epsfig}

\usepackage[psamsfonts]{amssymb}
\usepackage{amsfonts}\usepackage{amscd}
\usepackage{amsmath,amssymb}
\usepackage{cases}

\newcommand{\ba}{\begin{eqnarray}}
\newcommand{\ea}{\end{eqnarray}}
 \def  \Li3 { {\rm {Li}_3}  }
 \def  \li2  { { \rm {Li}_2 } }
\def  \poly4  { { \rm {Li}_4 } }
\def \Imm { \mbox{\rm Im} }

\usepackage{graphicx}\usepackage{epsfig}
\usepackage{amsfonts}\usepackage{amscd}
\usepackage{amsmath,amssymb}
\everymath{\displaystyle}
\usepackage[cp1251]{inputenc}
\usepackage[english,russian]{babel}

\begin{document}
\title{
 Contributions of QED diagrams
with vacuum polarization insertions to the lepton anomaly within the Mellin--Barnes representation}

\author{O.P. Solovtsova}
\email{olsol@theor.jinr.ru ;solovtsova@gstu.gomel.by}
\affiliation{Bogoliubov Lab.
Theor. Phys., JINR, Dubna,
141980, Russia}
 \affiliation{Gomel State Technical
University, Gomel, 246746, Belarus}
\author{V.I. Lashkevich}
\email{lashkevich@gstu.gomel.by}
\affiliation{Gomel State
Technical University, Gomel, 246746, Belarus}
\author{L.P. Kaptari}
\email{kaptari@theor.jinr.ru} \affiliation{Bogoliubov Lab.
Theor. Phys., JINR, Dubna,
141980, Russia}

\begin{abstract}
We investigate the radiative QED corrections to the lepton ($L=e,~\mu$
and $\tau$) anomalous magnetic moment arising  from vacuum
polarization diagrams by four closed lepton loops.
The method is based on the consecutive application of dispersion
relations for the polarization operator and the Mellin--Barnes
transform for the propagators of massive particles.
This allows one to obtain, for the first time, exact analytical
expressions for the radiative corrections to the anomalous magnetic
moments of leptons from diagrams with insertions of four identical
lepton loops all of the same type $\ell$ different from the external
one,  $L$. The result is expressed in terms of the mass ratio
$r=m_\ell/m_L$. We investigate the behaviour  of the exact analytical
expressions at $r\to 0$ and $r\to \infty$ and compare with the
corresponding asymptotic expansions known in the literature.
\end{abstract}

\noindent
\pacs{13.40.Em, 12.20.Ds, 14.60.Ef}

\maketitle

\thispagestyle{empty}

\section{INTRODUCTION}

Among the most important consequences of the Dirac theory is the
prediction~\cite{dirac} that the gyromagnetic factor $g_L$ of a lepton
$L$ ($L=e,~\mu$ and $\tau$) is $g_L=2$. However, the  self-interaction
with photons leads to a gyromagnetic factor $g_L\neq 2$, which in the
literature is referred to as the lepton anomaly, $a_L=(g_L-2)/2\neq
0$. Obviously, this anomaly is an important characteristic of the
magnetic field surrounding a lepton and, in spite of its extremely
small deviation from zero, it can serve as a substantial test of the
Standard Model (SM) or even  can indicate the  existence of some ``new
physics'' beyond the SM. Clearly, the self-energy correction to the
lepton electromagnetic vertex originates not only from the
electromagnetic interaction but also from strong and weak
interactions. A comprehensive review of contributions of different
mechanisms to $a_L$ can be found in, e.g.,
Refs.~\cite{Jegerlehner:2017gek,review-2021}. At present, experimental
measurements of $a_L$  for electrons~\cite{Parker:2018vye,Morel} and
muons~\cite{E989,Fermilab2023} are performed with an extremely high
accuracy  which imposes appropriate  requirements on  theoretical
calculations.

The first theoretical calculation of the leading order correction was
performed long ago by J.~S.~Schwinger~\cite{Schwinger1948} who showed
that $a_e=\alpha/2\pi$, where $\alpha$ is the fine structure constant.
Next to the leading order corrections involve much more diagrams which
result in complicate and cumbersome calculations.
Currently,  calculations of the eighth- and tenth-order quantum
electrodynamic (QED) corrections to $a_L$, which are important in
reduction of the theoretical uncertainties, are mainly  performed
numerically.
The corresponding calculations are rather computer resources consuming
(and require double checking, see, e.~g., \cite{Volkov}) and a
detailed study of the role of different mechanisms contributing to
$a_L$ are hindered. Therefore, it is enticing to find at least  a
subset of specific  Feynman diagrams which can provide analytical
expressions even if only for a restricted number of perturbative
terms. Then, having at hand analytical expressions, one can perform
calculations with any desired  accuracy and, consequently,  use  as
excellent tests of the reliability of direct numerical procedures. It
turns out that the subset of diagrams with loops originating only from
insertions of the photon polarization operator, the so-called
`bubble'-like diagrams, allows for analytical calculations of
corrections up to fairly high orders.
As known the Mellin--Barnes representation technique is widely used in
multi-loop calculations in high-energy physics,
c.f.~Refs.~\cite{Boos,Friot:2005cu,Kotikov:2018wxe,Mellin-book}.
As a first formulation of the approach based on Mellin--Barnes
integral representation as applied to calculations the lepton anomaly,
one can mention Ref.~\cite{Aguilar:2008qj}, where the corrections to
the muon anomaly of the eighth and tenth order (w.r.t. the
electromagnetic coupling constant $e$) were calculated in analytical
form as asymptotic expansions at mass ratio  $r=m_\ell/m_L \ll 1$.
Further generalization of the approach to obtain exact analytical
expressions for arbitrary $r$ ranging in the interval $0 < r <
\infty$, was reported in detail in Ref.~\cite{Solovtsova23}.
This paper can be considered as a continuation of our previous
research~\cite{Solovtsova23} of the bubble-diagram contributions to
$a_L$ using the Mellin-Barnes representation. Here we focus on the
exact analytical forms of the contributions to $a_L$ coming from
diagrams with insertions of four identical lepton loops.

\begin{figure}[t]
\begin{center}
\includegraphics[width=127mm]{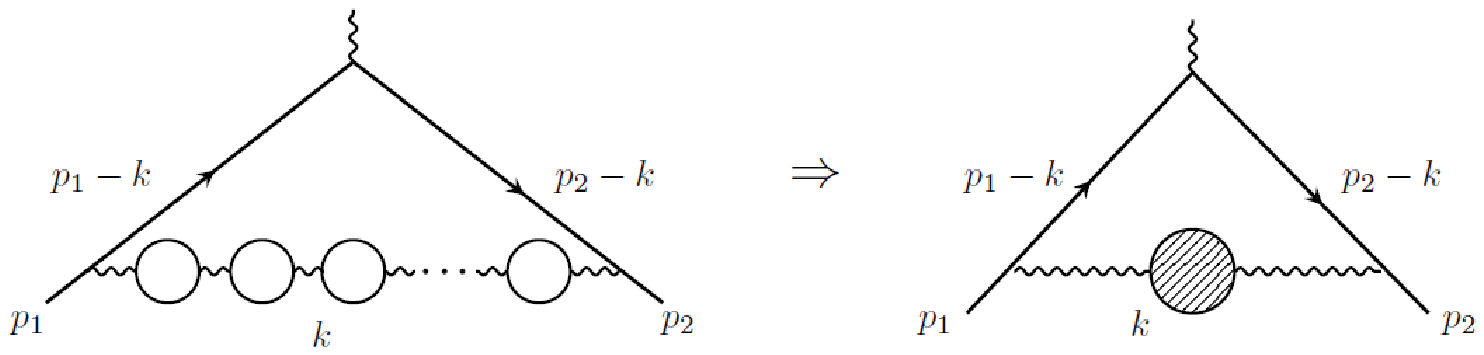}
\vspace{-1mm}
\caption{Left panel: radiative corrections to the electromagnetic lepton vertex
with insertions of the vacuum polarisation operator with an arbitrary number
of lepton loops. Right panel:
the second order diagram representing the set of graphs depicted in the left panel as exchanges of one massive
photon.}
\end{center}
 \label{Risodin}
\vspace{-5mm}
\end{figure}
\label{sec:1}
\section{THEORETICAL FRAMEWORK}

The main idea of the approach is to apply the dispersion relations to
the corresponding Feynman diagram to express it via the Feynman
$x$-paramet\-rization  of the second-order diagram with massive
photons and finally to apply the Mellins--Barnes representation to the
massive photon propagator (see Fig.~1 as the illustration) and again
the dispersion relations to the polarisation operators  of the
internal lepton $\ell\neq L$ different from the external one. In this
way, one can express any diagram from the mentioned subset in a rather
simple form as a comvolution integral of two Mellin momenta (for
details, cf. Ref.~\cite{Solovtsova23}). Then the QED corrections to
the lepton anomalous magnetic moment due to bubble-like Feynman
diagrams with the insertion of the photon polarization operator with
an arbitrary number $n=p+j$ of loops, where $p$ is the number of loops
formed by  leptons $L$ of the same type as the external one, $j$
denotes the leptons $\ell\neq L$, has the form
 \ba
 &&
  \hspace*{-8mm}
a_L(p,j)=\frac{\alpha}{\pi} \frac{F_{(p,j)}}{2\pi i}
\int\limits_{c-i\infty}^{c+i\infty} ds \;
 \left( \frac{4m_{\ell}^2}{m_L^2}\right)^{-s} \Gamma(s)\Gamma(1-s)\;
\left(\frac{\alpha}{\pi}\right)^{p}\Omega_p(s)
\left(\frac{\alpha}{\pi}\right)^{j}R_j(s), \nonumber
\\[-0.2cm]
&& \label{fin1} \ea where the factor $F_{(p,j)}$ is related to the
binomial coefficients $C_{p+j}^p$ as $F_{(p,j)} =(-1)^{p+j+1}
C_{p+j}^p$. Explicitly, the Mellin momenta $\Omega_p(s)$ and $R_j(s)$
read as
 \ba &&
\left(\frac{\alpha}{\pi}\right)^p\Omega_p(s)=
  \int_0^1 dx \; x^{2s} (1-x)^{1-s}
  \left[ \Pi^{(L)} \left( -\frac{x^2 }{1-x}m_L^2  \right)
  \right ]^p, \label{Omp1}
  \\[0.2cm] &&
\left(\frac{\alpha}{\pi}\right)^j R_j(s)= \int_0^\infty
\frac{dt}{t}\left( \frac{4m_{\ell}^2}{t}\right)^s \frac 1\pi \Imm
\left[ \Pi^{(\ell)} ( t)\right ]^j \label{rjj1}. \ea Below we consider
the case when $p=0$ and $j=4$ (see Fig.~2).

\label{sec:2}
\section*{Contribution of the diagram with four identical lepton loops}

\begin{figure}
    \begin{center}
        \includegraphics[width=0.33\textwidth]{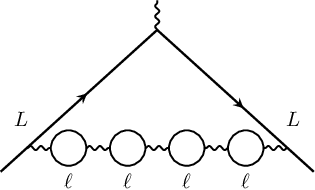}
        \caption{Vacuum polarization diagram  with
        insertion of four identical lepton loops formed by leptons other than the external one. }
    \end{center}
    \label{4loops}
    \vspace{-5mm}
\end{figure}

Using Eqs.~(\ref{fin1}) -- (\ref{rjj1}), the contribution to the
lepton anomaly from the diagram shown in Fig.~2 can be written as
\begin{equation}
a_{L}^{\ell\ell \ell  \ell}(r)  \equiv  A_{2,L}^{(10),\ell\ell \ell  \ell}(r) \left(\frac\alpha\pi\right)^5
= \left(\frac\alpha\pi\right)^5\frac{
1}{2\pi i}\int\limits_{c-i\infty}^{c+i\infty} r^{-2s}{\cal F} (s) ds\,
, \label{fin4loops}
\end{equation}
where the integrand  ${\cal F} (s)$ looks like
\begin{equation}
{\cal F} (s) = \left \{\frac{Z_1(s)}{729}-(1+s)Z_2(s)\left
[\frac{81\pi^2}{729}-\frac{2}{3}\;\psi^{(1)}(s) \right ] \right \}
\frac{ 4\pi^2 (1-s)}{Y(s)\sin^2(\pi s)} \, . \label{fs}
\end{equation}
%
For the sake of brevity the following notation has been introduced
\begin{eqnarray}
&& Z_1(s)=1259712+955332s-4110912s^2-6558755s^3-1384529s^4 \label{Z1}\\
&&
~ +3898617s^5 +3867513s^6+1653510s^7+373944s^8+43520s^9+2048s^{10} , \nonumber  \\
&& Z_2(s)=
-8400-26340s-22144s^2+1641s^3+11729s^4+6894s^5+  \nonumber \\
&& ~~~~~~~~  +1835s^6+237s^7+12s^8 , \label{Z2}
\\
&& Y(s)=  s(s+1)^2(s+2)^2(3+s)(4+s)(5+s) (6+s)(1+2s)(3+2s)
 \nonumber\\
&& ~~~~~~ \times   (5+2s)(7+2s). \label{YZ}
\end{eqnarray}

As the  integrand (\ref{fs}) is singular,  then the integral
(\ref{fin4loops}) can be carried out by the Cauchy residue theorem
closing the integration contour in the left ($r<1$) or   right ($r>1$)
semiplanes of the Mellin complex variable $s$ and computing the
corresponding residues in these domains.

{\underline{Case $r>1$.}} By closing the contour of integration to
the right and computing the corresponding residues in this domain,  we
get the following result:
\begin{eqnarray}  \label{A10B}
&&  \hspace{-5mm} A_{2,L}^{(10),\ell\ell \ell  \ell}(r>1)
  = \;
D_0(r) + 2 D_1(r) \ln (r) + D_2(r) \bigg[{\rm
Li}_2\left(\frac{1}{r^2}\right) -2\ln \left(1-{\frac{1}{r^2}}\right)
\nonumber\\[0.2cm]
&&  \hspace{-8mm}  \times \ln (r) \bigg] -\frac{16}{9}\left(\frac{8
r^2}{75}+5 r^4+\frac{2 \pi ^2 r^4}{3}\right)
 \left[ {\rm
Li}_2\left(\frac{1}{r^2}\right) \ln (r)+ {\rm
Li}_3\left(\frac{1}{r^2}\right)\right]  + \Sigma_1(r),
\end{eqnarray}
where Li$_n$ is the polylogarithm function of the order $n$ and the
polynomials $D_{0-2}(r)$ are defined as follows
\begin{eqnarray}
&& \hspace{-5mm} D_0(r) =
\;-\frac{14463825527}{11252115000}-\frac{143175013 r^2}{14033250}+
\frac{680597537 r^4}{63149625}+\frac{97213348
r^6}{63149625}+\frac{797842 r^8}{2338875} \quad
\nonumber\\[0.2cm]
 && ~ +\frac{42952 r^{10}}{1002375}
+\frac{~\pi^2 }{2}r \left(\frac{18203}{31185} +\frac{28010
r^2}{5103}-\frac{5957 r^4}{2025}-\frac{12916 r^6}{70875}\right) -\pi^2
\left(\frac{57419}{255150} \right.
\nonumber\\[0.1cm]
&& \hspace{-5mm} \left. -\frac{2878}{2679075 r^2}-\frac{1251149
r^2}{1559250}-\frac{249589 r^4}{841995}
 -\frac{40591 r^6}{336798}-\frac{3632
r^8}{93555}-\frac{64 r^{10}}{13365}\right)
+2r\bigg[\frac{18203}{31185}
\nonumber\\[0.2cm]
&& \hspace{-5mm} +\frac{28010 r^2}{5103}-\frac{5957 r^4}{2025}
 -\frac{12916 r^6}{70875}
-\pi ^2 \left(\frac{18203}{748440 }+\frac{2801r^2}{13608
}-\frac{23r^4}{216} -\frac{7r^6}{1080} \right) \bigg]
\nonumber\\
&& \hspace{-5mm} \times \bigg[{\rm Li_2} \left(\frac{1-r}{1+r}
\right)-{\rm Li_2}\left(-\frac{1-r}{1+r}\right) \bigg]
+\frac{~\pi^4}{2} r \left(\frac{18203}{748440 }+\frac{2801 r^2}{13608
}-\frac{23 r^4}{216} -\frac{~7 r^6}{1080} \right), \nonumber
\end{eqnarray}

\begin{eqnarray}
&&  D_1(r)= \;\frac{9239297}{26790750}+\frac{15360524
r^2}{5011875}-\frac{109392281 r^4}{21049875}-\frac{26671558
r^6}{21049875}-\frac{2468692 r^8}{7016625}
\nonumber\\[3mm]
 && -\frac{42952r^{10}}{1002375} +\pi ^2 \left(\frac{25}{243}-\frac{1185953 r^2}{1871100}-\frac{4624 r^4}{40095}
-\frac{26501 r^6}{224532}-\frac{416 r^8}{10395}-\frac{64
r^{10}}{13365}\right), \nonumber
\\[3mm]
&& D_2(r) = \;\frac{8}{175}+\frac{13664 r^2}{10125}+\frac{1274
r^4}{729} -\frac{1568 r^6}{1215}+\frac{410 r^8}{567}+\frac{42152
r^{10}}{127575}+\frac{42952 r^{12}}{1002375}
\nonumber\\
 && ~~~~~~~~~ +\pi ^2 \left(\frac{2}{~81}-\frac{8 r^2}{27}+\frac{34 r^4}{81}-\frac{64 r^6}{405}+\frac{16 r^8}{189}+\frac{~~64
r^{10}}{1701}+\frac{64 r^{12}}{13365}\right).\nonumber
\end{eqnarray}

Finally, the last term in Eq.~(\ref{A10B}) is the sum associated with
 $\sin ^2(\pi s)$ in the denominator of the function ${\cal F} (s)$:
\begin{equation}\label{4-loop-SumB}  
\Sigma_1(r)=\; \frac{8}{3}
\sum_{n=2}^{\infty}\left[\frac{C_1(n)}{Y(n)}\psi_n^{(1)}+( n-1)
C_2(n)\left(2\psi_n^{(1)}\ln(r)-\psi_n^{(2)}\right)\right]
\frac{r^{-2n}}{Y(n)} \,,
\end{equation}
where
\begin{eqnarray}
  && C_1(n)=\left(n+1\right)^2 (n+2)\bigg(635040000+7687008000 n+36734547600n^2
 \nonumber\\
 && +93125888040n^3 +135651027372n^4 +104915891978n^5+10006706560n^6
  \nonumber\\
 &&-69851951805n^7-83164962406n^8  -51439049641n^9-18649902420n^{10} \nonumber\\
 &&
-2892341259n^{11}+812142446n^{12}+656337939n^{13}  +212614912n^{14}\nonumber\\
  &&
+42833116n^{15}+5711184n^{16}+493456n^{17}+25152n^{18}+576n^{19}\bigg)\,
,
 \label{C1}
\end{eqnarray}
$C_2(n)=Z_2(s=n)$, $Z_2(s)$ and $Y(s)$ are given  by Eqs.~(\ref{Z2})
and (\ref{YZ}), respectively; $\psi_n^{(1,2)}$ denotes the polygamma
functions of the first or the second order of the integer argument
$n$.

{\underline{Case $r<1$.}} It should be noted that calculating the
integral (\ref{fin4loops}) in the region $r<1$ is much more difficult
compared to the case $r>1$.
This is due to the presence of additional zeros for the function
$Y(s)$, Eq.~(\ref{YZ}). Also, for negative arguments, the polygamma
function $\psi^{(1)}(s)$ also has poles for integers $s=-n$.
Having found all the residues and summed them up, we get
\begin{eqnarray}  \label{A10A}
&& A_{2,L}^{(10),\ell\ell \ell  \ell}(r<1) =  P_0(r) + 2P_1(r) \ln (r) +
4P_2(r)\ln^2 (r)+8P_3(r)\ln^3 (r)
\nonumber\\[0.2cm]
& &+\frac43 K_1\ln^4(r) - \frac{128}{135}r^4\ln^5(r)+2 K_3\bigg[\Phi
\left(r^2,4,\frac{1}{2}\right)- 2\Phi
\left(r^2,3,\frac{1}{2}\right)\ln (r)
\nonumber\\
&&  + 2\Phi \left(r^2,2,\frac{1}{2}\right)\ln^2 (r)\bigg]+
\Sigma_2(r),
\end{eqnarray}
where $\Phi(r^2,n,1/2)$ is the Lerch function, which is related to the
polylogarithms as: $ \Phi(r^2,n,1/2)={2^{n-1}}\bigl[{\rm Li}_n
(r)-{\rm Li}_n (-r)\bigr]/{r}. $
The notation $P_i(r)$ corresponds to the expressions
\begin{eqnarray}
&& P_0(r)= \frac{64613}{26244}-\frac{145231
r^2}{17325}+\frac{265354583 r^4}{7016625} +\frac{5155111
r^6}{1002375}+\frac{1644584209 r^8}{261954000}
\nonumber\\[0.2cm]
&& +\frac{262864711931 r^{10}}{445583754000}-\frac{38750851857953
r^{12}}{70020304200000} + \pi^2 \left(\frac{317}{1458}-\frac{380911
r^2}{173250} \right.
\nonumber \\
[0.2cm]
 &&
 \left.
+\frac{1224743 r^4}{841995}-\frac{1473151 r^6}{1559250}-\frac{9577847
r^8}{61122600}
+\frac{~283177187 r^{10}}{3713197950}\right.\nonumber\\[0.1cm]
 &&
 \left.
 +\frac{~55905529021
r^{12}}{1283705577000}\right) -\pi^2 K_1(r)\left({\rm Li}_2(r^2)
-\frac{\pi^2}{5} \right)+\frac{\pi^2}{2r}\left( K_4(r)-
\frac{\pi^2}{3}K_3(r)\right)
\nonumber\\
 &&
-\frac{2}{r}\left(K_4(r)+\pi ^2 K_3(r)\right)\left[{\rm
Li}_2\left(\frac{1-r}{1+r}\right)-{\rm
Li}_2\left(-\frac{1-r}{1+r}\right)\right]-K_5(r){\rm Li}_2 (1-r^2)
\nonumber\\
 &&
 -\left(\frac{100}{81}+\frac{128 r^2}{675}+\frac{64 r^4}{9}- \pi^2
\frac{~32 r^4}{9}\right){\rm Li}_3 (r^2) +\left(6
K_2(r)-\frac{128r^4}{675} \right.
\nonumber\\
 &&
 \left.
 - \pi^2\frac{~64r^4}{27}\right)\zeta(3)
-2\left( K_1(r)+\frac{8}{27}\right){\rm Li}_4 (r^2)+\frac{128}{9}r^4
\left({\rm Li}_5(r^2)-\zeta(5)\right) , \nonumber
\end{eqnarray}
\begin{eqnarray}
&& P_1(r)= \frac{8609}{4374}-\frac{2190631
r^2}{280665}-\frac{~139188328 r^4}{7016625} -\frac{1590044
r^6}{7016625}-\frac{27882949 r^8}{9355500}
\nonumber\\
 && -\frac{1090421197
r^{10}}{1768189500}+ \frac{671218651 r^{12}}{5051970000}+\pi^2
 \left(\frac{50}{243}-\frac{50399 r^2}{41580}+\frac{6431 r^4}{13365}
\right.
\nonumber\\
&&  \left.  +\frac{20003 r^6}{1871100}+\frac{5539 r^8}{48510}
-\frac{16846 r^{10}}{5893965} - \frac{2119877 r^{12}}{92619450}\right)
 +\left(\frac{50}{81}+\frac{64 r^2}{675}+\frac{32
r^4}{9} \right.
  \nonumber\\
&& \left. -\pi^2 \frac{16 r^4}{9}\right){\rm Li}_2 (r^2)-\pi^2
K_1(r)\ln(1-r^2) +  2 K_1(r)\biggl({\rm Li}_3 (r^2) +2\zeta(3)\biggr)
  \nonumber\\
&& + \;\frac{4}{27}\;{\rm Li}_3(r^2)-\frac{~32r^4}{3}{\rm Li}_4 (r^2)
, \nonumber
 \end{eqnarray}
 \begin{eqnarray}
P_2(r) &=& \;\frac{317}{486}-\frac{247663 r^2}{62370}+\frac{2139227
r^4}{561330} -\frac{40591 r^6}{112266}-\frac{3632
r^{8}}{31185}-\frac{64 r^{10}}{4455}
\nonumber\\
 &&
+ K_1(r)\left({\rm Li}_2 (r^2)-\frac{\pi^2}{3}\right) +\frac{32}{9}r^4
\biggl({\rm Li}_3 (r^2)-\zeta(3)\biggr) , \nonumber
\end{eqnarray}
\begin{eqnarray}
P_3 &=& \frac{25}{243}-\frac{231277 r^2}{374220}-\frac{4624
r^4}{40095}-\frac{26501 r^6}{224532}-\frac{416 r^8}{10395}-\frac{64
r^{10}}{13365}-\pi^2\frac{~16 r^4}{81}
\nonumber\\[0.1cm]
 && -\frac{16}{27} r^4\; {\rm Li}_2
(r^2)-\frac13K_1(r) \ln (1-r^2) - K_3(r)
\frac{1}{3r}\bigl[\ln(1+r)-\ln(1-r)\bigr], \nonumber
\end{eqnarray}
where $\zeta(x)$ denotes the Euler--Riemann zeta function of the
argument $x$.

\vskip 0.1cm

Finally, the sum $\Sigma_2 (r)$ in Eq.~(\ref{A10A}) reads as
\begin{equation}  
\Sigma_2 (r)= \frac{8}{3} \sum_{n=7}^{\infty}\left[\frac{~
C_1(-n)}{Y(-n)}\psi_n^{(1)}+(n^2-1)
Z_2(-n)\left(2\psi_n^{(1)}\ln(r)+\psi_n^{(2)} \right)\right]
\frac{r^{2n}}{Y(-n)}, \label{4-loop-SumA}
\end{equation}
where the notations are the same as in the sum (\ref{4-loop-SumB}).
For brevity, polynomials $K_i(r)$ in Eq.~(\ref{A10A}) are introduced
\begin{eqnarray}
 &&\hspace{-5mm}  K_1(r)=\frac{2}{27}-\frac{8 r^2}{9}+\frac{34 r^4}{27}
 -\frac{64 r^6}{135}+\frac{16 r^8}{63}+\frac{64 r^{10}}{567}
+\frac{64 r^{12}}{4455}\,,
\nonumber \\
[0.2cm]
 &&
\hspace{-5mm} K_2(r)=\frac{50}{243}-\frac{104 r^2}{81}-\frac{134
r^4}{81}-\frac{4432 r^6}{6075}-\frac{1033 r^8}{2205}-\frac{12332
r^{10}}{535815}+\frac{2119877 r^{12}}{46309725}\,,
\nonumber \\
 &&
\hspace{-5mm} K_3(r)=\frac{18203r^2}{249480}+\frac{2801
r^4}{4536}-\frac{23 r^6}{6075}-\frac{7 r^8}{360}\,,
\nonumber \\
 &&
\hspace{-5mm} K_4(r)=-\frac{~2801 r^4}{567}+\frac{230
r^6}{81}+\frac{1813 r^8}{10125}\,,
\nonumber \\
 && \hspace{-5mm}
K_5(r)=-\frac{317}{243}+\frac{31664 r^2}{10125}+\frac{167
r^4}{729}-\frac{32 r^6}{27} +\frac{56 r^8}{81}+\frac{1640
r^{10}}{5103}+\frac{3832 r^{12}}{91125}\,. \nonumber
\end{eqnarray}

\label{sec:3}
\section{DISCUSSION}

The above Eqs.~(\ref{A10B}) and (\ref{A10A}) represent the exact
analytical expressions of the tenth order of the radiative corrections
from diagrams with the insertion of four identical lepton loops, as
depicted in Fig.~2. Despite their cumbersomeness, the explicit
analytical  form allows for numerical calculations with any desired
precision. The precision can only be limited by the knowledge of the
basic physical quantities such as $\alpha$, $m_\ell$ and $m_L$.

\begin{figure}
    \begin{center}
     \includegraphics[width=0.95\textwidth]{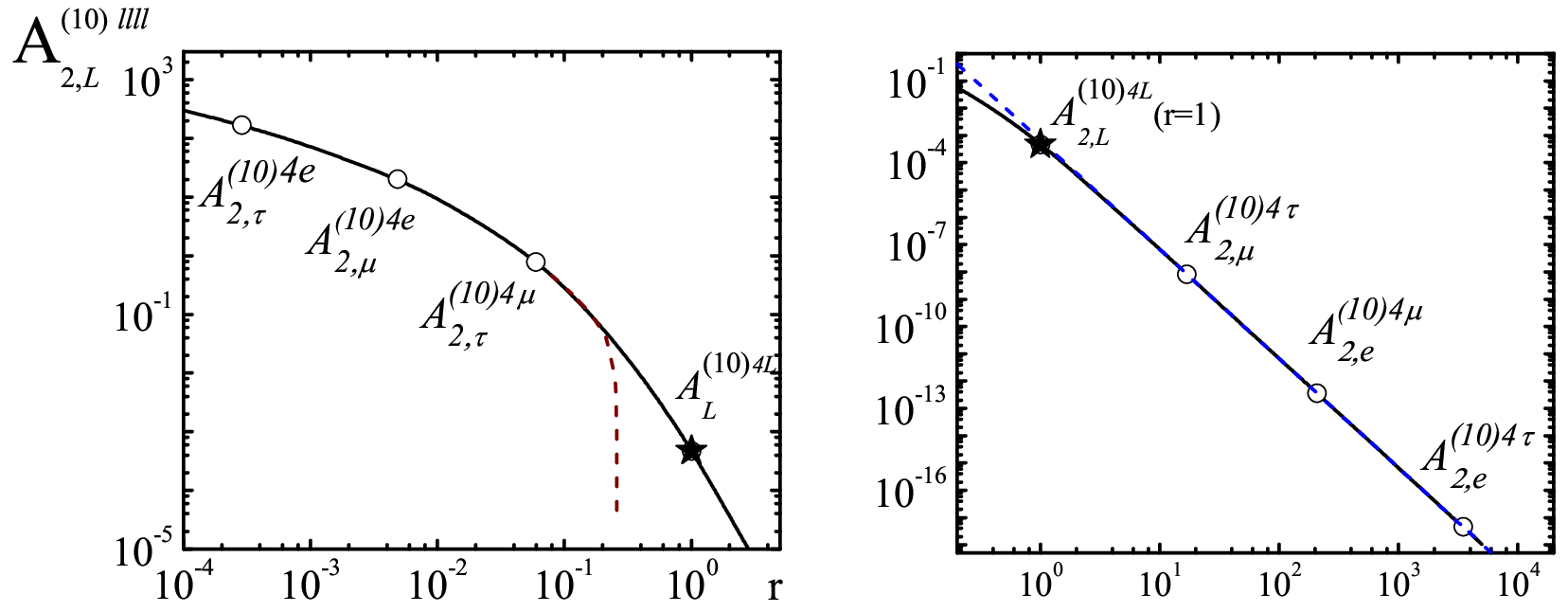}
        \caption{Comparison of asymptotic expansions
          with exact calculations of the coefficient $A_{2}^{(10)}(r)$
        The solid curve is the exact result, the dashed  lines are
        the asymptotic for $r\ll 1$ (left panel)
        and $r\gg 1$ (right panel). The open circles, as well as the labels associated with them,   point to
physical values of the ratio $r$ and to the corresponding physical
coefficients $A_{2,L}^{(10)\ell\ell \ell\ell}(r)$. The universal value
$A_{2,L}^{(10)LLLL}(r)$  is displayed by the full star.}
    \end{center}
 \label{RisAsymp}
    \vspace{-5mm}
\end{figure}

Let us briefly discuss asymptotic expansions of the coefficient
$A_{2}^{(10),\ell\ell \ell  \ell}$ considering the limits  $r\ll 1$ and
$r\gg 1$.
A comparison of the result of calculating $A_{2}^{(10),\ell\ell \ell  \ell}$  using
asymptotic and exact formulas,  Eqs.~(\ref{A10B}) and (\ref{A10A}), is
illustrated in Fig.~3, in which the solid curve is the exact result,
and the dashed-dotted line is the asymptotic for $r\ll 1$ (left
panel) and $r\gg 1$ (right panel). Note, we do not give here our
expansion formula for the case $r\ll 1$ (see discussion below). For
the case $r\gg 1$, using Eq.~(\ref{A10B}), we get
\begin{eqnarray}\label{asympR>1}
A_{2,L}^{(10),\ell\ell \ell  \ell}(r\gg 1) &\simeq& \;\left(
-\frac{369904}{88409475}+\frac{4402}{109147}\;\zeta(3)\right)\frac{1}{~r^4}-
\left(\frac{598587203}{82751268600} \right. \nonumber \\
&& \left. ~~~~~~~~  ~~~~~~~~  -\frac{71960}{127702575}\zeta(3) \right)
\frac{1}{~r^6} + {\cal {O}}\left(\frac{1}{r^8}\right)\,.
\end{eqnarray}
It is interesting to note that there are no logarithmic terms in this
expration.

One can see from Fig.~3 that the approximate expansions  practically
coincide with the exact formulae in quite large intervals of $r$,
namely $0<r<0.2$ for the expansion $r\ll 1$ (left panel) and $2<r<
\infty$ for the expansion $r\gg 1$ (right panel), herewith both
intervals include all the corresponding physical values of
$a_L^{\ell\ell\ell\ell}$.

Let us return to the region $r \ll 1$ for which there are asymptotic
expansions in the literature,  see
Refs.~\cite{Aguilar:2008qj,Laporta94}.
Our asymptotic expansion completely coincides with the expansion given
in Ref.~\cite{Laporta94}, which, however, corresponds only to the
order $O(r^2)$.
In the expansion Ref.~\cite{Aguilar:2008qj}, see Eq.~(A9) in the
Appendix, which is given up to the order $O(r^5)$, we found two
misprint: a different sign in the term $\frac{32}{27}r^2 \ln^4(r)$ and
in $-\frac{61}{27}\pi^2 \zeta(5) r^4$ instead of 61 there is the
number 64.

Thus, the present investigation has confirmed that the approach used here is a
powerful tool for finding exact analytical expressions for the
bubble-like diagrams contributions to the anomalous magnetic moment of leptons.

\section*{Acknowledgments}
This work was supported in part by a grant
of the JINR--Belarus collaborative program.

\end{document}